\documentclass[12pt]{iopart}
\usepackage{graphicx,iopams,setstack,cite}
\begin{document}

\title[Confinement effects on monodisperse disk packings]{Confinement effects on the configurational order of monodisperse disk packings}

\author{Bj\"orn Arnold$^1$, Ayse Turak$^2$\footnote{Permanent address: Department of Engineering Physics, McMaster University, Hamilton, Ontario, Canada.}, and Alejandro D\'{\i}az Ortiz$^1$}
\address{$^1$INTRON GmbH, D-74523  Schw\"abisch Hall, Germany, EU}
\address{$^2$Faculty of Engineering and Natural Sciences, Sabanci University, 34956 Istanbul, Turkey} 

\ead{alejandro.diazortiz@gmail.com}

\date{\today}

\begin{abstract}
Monodisperse circular disks have been collectively packed in confined geometries using a Monte Carlo method where the compaction is propelled by two-dimensional stochastic agitation. We have found that  confinement  (i.e., finite-size plus surface effects) determines the symmetry of the packed configurations together with the size evolution of the probability density of the packing fraction. For the particular case of small systems  in square containers, the probability density of the packing fraction shows several well-defined peaks, depending on the system size, for both hard-wall and periodic boundary conditions. High-symmetry configurations (other than the $n$$\times$$n$ square arrays) with non-negligible occurrence probabilities are found as a direct consequence of monodispersity and confinement.
\end{abstract}

\submitto{\JPCM}
\maketitle

\section{Introduction}
\label{sec:intro}
Confined monodisperse systems have recently attracted significant attention for their potential in technological applications and their role in explaining fundamental questions on self organization\cite{langevin_monodisperse_foams_2010,whitesides_microfluidics_2006,hoehler_bubbles_2011,shen_confinement_feature_2010,meaking_confinement_channels_2001,kumacheva_confinement_channels_2003,gates_monodisperse_shape_size_2001,ozin_confinement_patterning_2001,kumacheva_confinement_patterning_2002}. For instance, current experimental techniques are able to produce extremely monodisperse foams, bubbles, and colloidal nanoparticle suspensions that are stable against coalescence and coarsening in a time span of several days.  This renaissance in monodisperse systems has been spurred by interest in manipulating fluids in micrometric channels (microfluidics)\cite{langevin_monodisperse_foams_2010}. The possible applications of microfluidics range from chemical and biological analysis to optics to information technologies\cite{whitesides_microfluidics_2006}. In addition, there are striking analogies between monodisperse bubbles and granular materials that extend to both jamming and flow\cite{hoehler_bubbles_2011}. 

Monodisperse systems tend to self organize in nontrivial ways under confinement\cite{shen_confinement_feature_2010}. Notably, monodisperse colloidal particles have shown a rich configurational landscape in two-dimensional channels\cite{meaking_confinement_channels_2001,kumacheva_confinement_channels_2003} and planar patterning on surfaces\cite{gates_monodisperse_shape_size_2001,ozin_confinement_patterning_2001,kumacheva_confinement_patterning_2002}. The possibility of controlling the state of order in monodisperse particles by confinement has applications ranging from photonic crystals to drug-delivery  to sensor technologies\cite{shen_confinement_feature_2010}.

In the cases discussed above, different mechanisms were employed to activate and sustain the self-assembly process. For example, a laminar flow was used to stabilize the novel crystal structures of colloidal particles in microchannels in Ref.~\cite{kumacheva_confinement_channels_2003}, whereas an electrostatic field aided the self-assembly of electrodeposited particles of Ref.~\cite{kumacheva_confinement_patterning_2002}. A very attractive and viable process for self-assembly is vibration, either in the form of thermal or mechanical agitation. The tapping experiments in granular materials are well-known examples of the latter\cite{pursuit_perfect_packing_2008}. Vibration is at the core of the self-assembly of micro-to-millimeter objects with a variety of functions from computing devices\cite{rothemund_agitation_2000}  to semiconducting devices (LEDs)\cite{whitesides_agitation_2002} to three-dimensional electric circuits\cite{whitesides_agitation_2005}. Quite recently, a high-yield vibration-based method to assemble magnetically interacting cubes has been reported\cite{whitesides_agitation_2011}.

Here we propose, and analyze the consequences of, a simple model for the self-assembly of two-dimensional confined hard disks under stochastic vibration (Sec.\ \ref{sec:model}). In contrast to previous works on confined systems of monodisperse disks or spheres, which have mostly been concerned with the random close packing limit (see, e.g., Ref.~\cite{weeks_confinement_2009} for a recent and excellent survey of the relevant literature), our goal here is to investigate the impact of the boundary conditions on the distribution of configuration states. Special emphasis is given to identifying nontrivial high-symmetry configurations and their yield rates---a subject of interest for applications on devices whose functionality is based upon different states of order\cite{shen_confinement_feature_2010,kumacheva_confinement_channels_2003,gates_monodisperse_shape_size_2001,ozin_confinement_patterning_2001,kumacheva_confinement_patterning_2002} (Sec.\ \ref{sec:results}).  

\section{Model approach and compaction protocol}
\label{sec:model}

We consider the following scenario:\ Two-dimensional monodisperse disks are distributed randomly, following a uniform distribution, in a square container with hard walls. The density of the system (initially diluted) is uniformly increased by compressing the container (or equivalently by expanding the disks) with a slow yet monotonic rate $\gamma$. During this compaction process, the particles are continuously agitated in the horizontal plane with frequency $\omega$ and amplitude $\bi{A}_i$. Each particle $i$ is individually and randomly shaken. The shaking amplitude is random, following a Gaussian distribution characterized by zero mean and standard deviation $\sigma_l$. The direction of the agitation is also random but, in contrast to the amplitude, the shaking direction follows a flat random angular distribution over $2\pi$. These choices are in place to guarantee that the particles sample different positions within the container with no preferential direction, i.e., a truly two-dimensional stochastic sampling. The compaction process continues, resolving particles overlaps through the stochastic agitation, until no further increase of the density is possible. A protocol based on this approach consistently appears to unbiasedly survey the configurational space, retrieving rare events with small, yet finite probability (see below).

We have implemented our model in a Monte Carlo approach where the particles are frictionless and represented by regular polygons\footnote{Note that a circle can be considered as a regular polygon with an infinite number of sides. In practice, however, we have found that polygons with as few as 100 sides produced results indistinguishable from those using (mathematical) circles.}. We assume a frictionless hard-body interaction between the particles and the walls of the container. The frequency of vibration $\omega$ becomes the number of shakes per particle. We have scanned the parameter space for the least-biased set of values for the growth rate $\gamma$, shaking frequency $\omega$, and $\sigma_l$\cite{ba_thesis_2009}---note that least-biased in this context means the subset of values where the error bars were minimized for at least 5 runs, and the packing fraction had reached a steady-state value with respect to the variable. We have found that a compaction protocol with $\gamma$=$10^{-5}$, $\omega$=1000, and $\sigma_l / d$=1 ($d$ is the disk diameter), offers a reasonable sampling of the configurational space without favoring particular disks arrangements. 

The significant length scales are the diameter of the disks and the size of the container. In an inflationary packing scheme, as the one used here, the ratio of those length scales becomes equivalent to  the total number of particles (see, for example, Ref.~\cite{kendall_confinement_vibration_2006}). In the following, we investigate how $n^2$ (=9--1024) disks are packed in square containers. Doing so allows us to pinpoint the limit where the confinement-induced ordering (with zero configurational entropy) gives way to disordered packings (with nonzero configurational entropy) as the number of particles increases. Specifically, we want to answer the following questions:\ How often $n^2$ disks crystallize in $n$$\times$$n$ square arrays with packing fraction $\pi/4$, when packed under the above-described protocol? How does the transition from square to (mostly) triangular symmetry occur? How do the confinement boundary effects fade in the large $n$ limit?

%
% Figure 1: Summary of results and comparison between HBC and PBCs.
%
\begin{figure}[t]
\centerline{\includegraphics[width=0.8\textwidth]{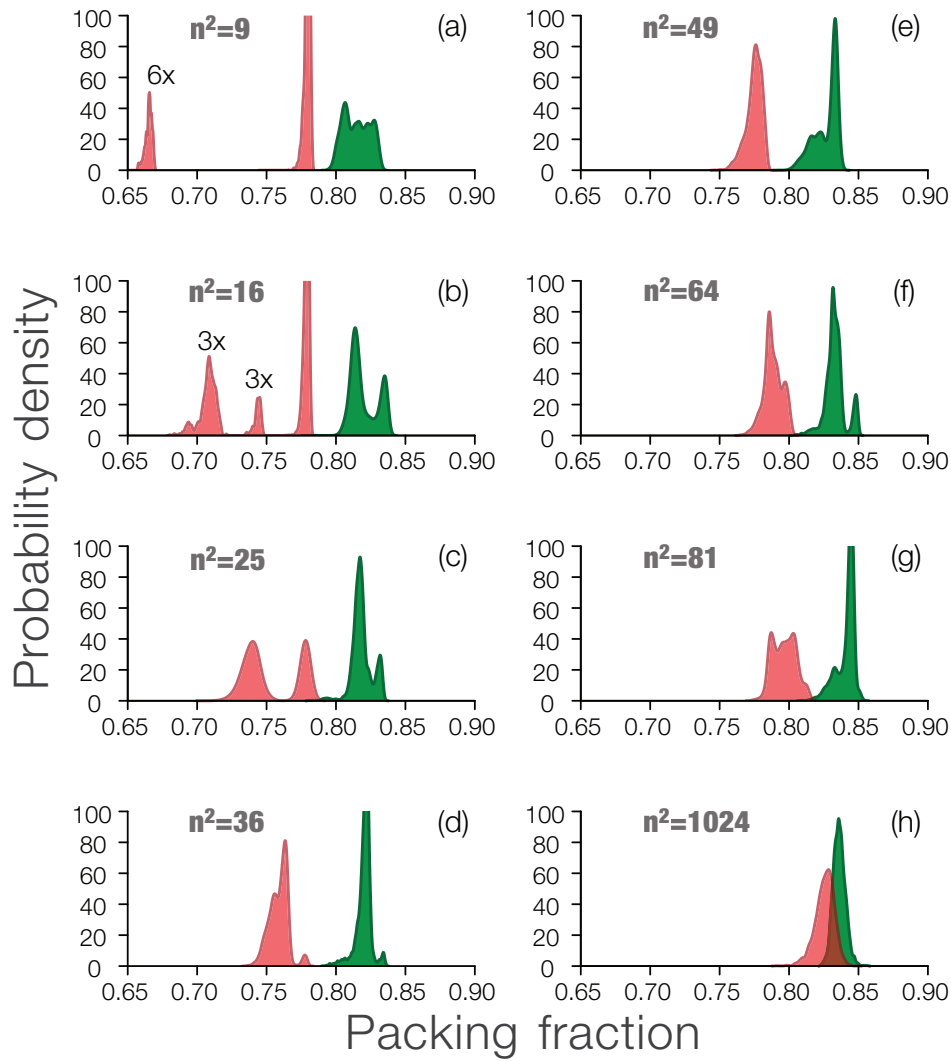}}%
\caption{\label{fig:pdfs_bcs} (Color online) Probability density  function for the packing fraction of $n^2$ particles in square containers with hard (red or light grey) and  periodic (green or dark grey) boundaries. (a)--(g)  $n^2$=9--81, respectively. The low-coverage peaks for the probability density in (a) and (b) have been magnified 6x and 3x, respectively, to facilitate their inspection. The additional structure (peaks) in the probability density functions are associated to special particle configurations for small ($n^2$$<$49) confined systems. For comparison, in (h) we show the PDFs for large systems with $n^2$=1024.}
\end{figure}

\section{Results}
\label{sec:results}
Figure~\ref{fig:pdfs_bcs} summarizes our findings for the probability density function (PDF) of the packing fraction for a series of $n^2$ number of particles in square containers. In all cases, we have built the PDF upon 4000 independent packed configurations (full compression runs), after assessing in selected cases that a higher sampling did not change the trends seen here. The most prominent feature displayed by PDFs is the number of peaks, with up to four distinct high symmetry configurations visible with finite probability. This is most apparent for systems with few particles ($n^2$$<$49) in hard-wall containers (red or light grey). It is also important to note the lack of any non-symmetric disordered packings with packing fractions between those visible peaks visible. The peak at packing fraction $\Phi$=$\pi/4$$\sim$0.78 corresponds to $n$$\times$$n$ square-array configurations.\footnote{Our numerical calculations of the packing fraction have an accuracy of $\pm6$\% for the containers with 9 particles. This is characteristic of small systems, with the efficiency of an inflationary algorithm increasing steadily with the number of particles. The conclusions of our study are independent and thus unaffected by the numerical precision of our calculations}   For containers with hard-walls (or hard-boundary conditions, HBCs), such square arrays dominate the configurational landscape for up to 25 particles, where the second peak observed around $\Phi$$\sim$0.74 also becomes significant. Square-array configurations become increasingly less probable as the system size increases; for $n^2$=36 only 3.7\% of the samples pack in such configuration. For $n^2$$\geq$49,  $n$$\times$$n$ square-array configurations have vanishing frequency probabilities, and disorded packings are increasingly dominant. After the relaxation of confinement effects which strongly encourage crystallization, the mean packing fraction increases with the number of particles as previously observed for polydisperse systems\cite{ohern_disks_2005,weeks_confinement_2009}. The shape of the distribution also changes, increasing in width as more random configurations begin to dominate. The occurrence of high symmetry configurations other than $n$$\times$$n$ is notable, even for systems as small as 16 particles, where 22\% of the configurations are different from the most probable. This number increases to 62\% for 25 particles in a square container with hard walls.

%
% Figure 2: Selected configurations for HBCs.
%
\begin{figure}[t]
\centerline{\includegraphics[width=0.8\textwidth]{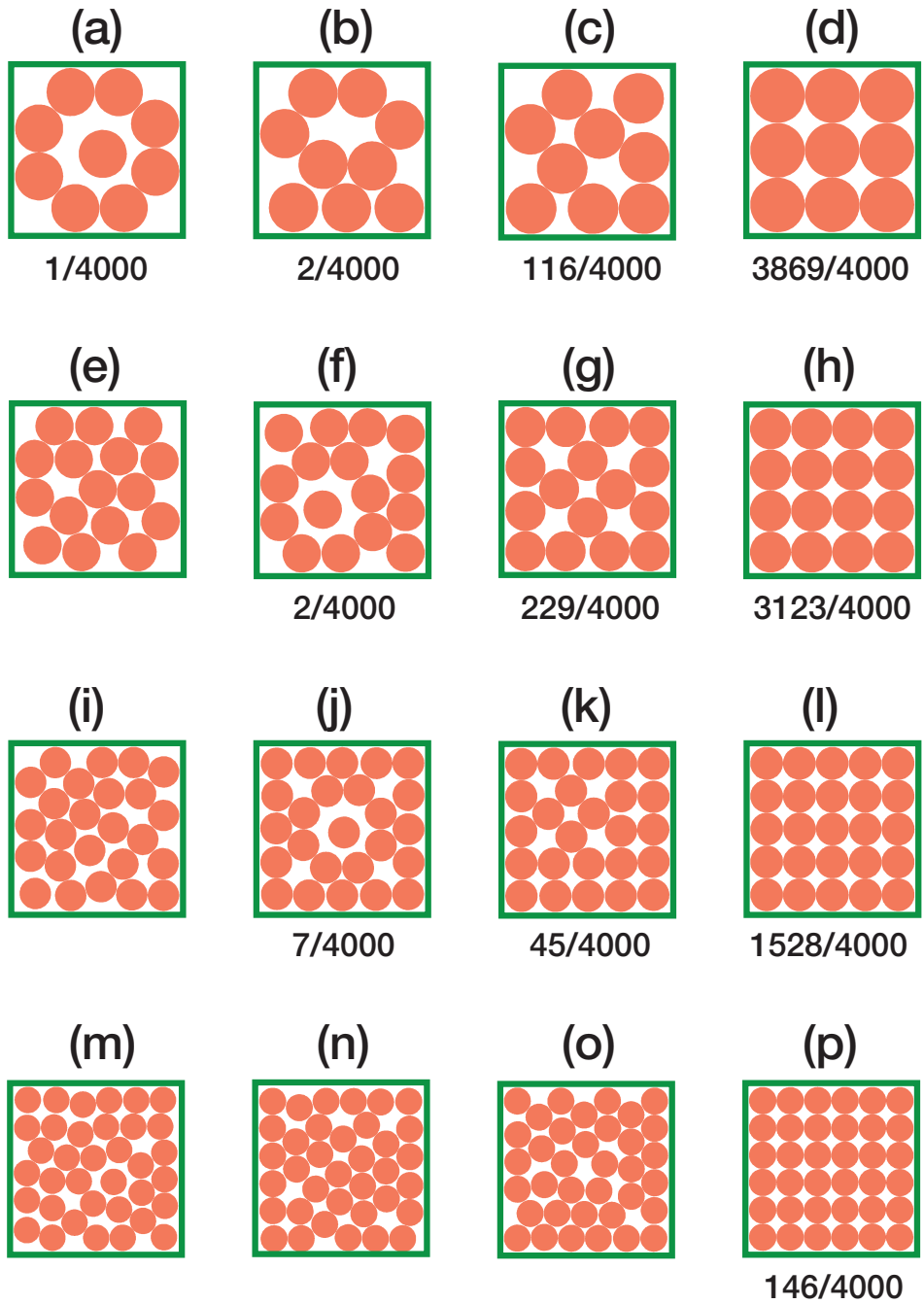}}%
\caption{\label{fig:selected_configurations} (Color online) Selected configurations for 9 [(a)--(d)], 16 [(e)--(h)], 25 [(i)--(l)], and 36 [(m)--(p)]  particles in a square container with hard walls. For the highly symmetric configurations, their sampling frequency is shown. The first column from the left contains the configuration with the lowest packing fraction, whereas those of the last column correspond to the highest packing ones.  Configurations (n) and (o) represent two typical 36-particle configurations around the peak of the PDF. Under the current protocol, our model did not retrieve high symmetry configurations for $n^2$$\ge$49.}
\end{figure}

The HBC features discussed above are the result of confinement, that is the combination of finite size plus surface effects. By contrast, due to the lack of surface effects, the PBC probability density functions show characteristics that are solely due to the system's finite size. A comparison between these two set of results allows therefore a separation of the two often conflated effects\cite {weeks_confinement_2009}.

For example, square-array configurations (with packing fraction $\pi/4$) are almost never observed for particles with periodic boundary conditions (PBCs). Instead, configurations with patches of square and triangular symmetry are favored for small systems with PBCs, leading to a bimodal structure in the probability distributions at low $n$. Similarly to O'Hern and co-workers\cite{ohern_disks_2005,ohern_disks_2006}, who used a completely different model and protocol,  we have also observed that even-odd effects are present in the shape of the distributions for systems with PBCs.

For larger systems with HBCs, the PDFs become single peaked. We can extrapolate the mean value of the packing fraction distribution $\Phi$ as a function of the number of particles $N$ in the container using\cite{weeks_confinement_2009,kendall_confinement_vibration_2006}
\begin{equation}
\Phi(N)=\Phi_{\rm bulk} +a N^{-1} ,
\end{equation}
with $\Phi_{\rm bulk}$=0.831 and $a$=$-2.82$. Notice that this value of $\Phi_{\rm bulk}$ is in the range of the corresponding random-close packing limit of 2D frictionless disks\cite{berryman_rcp_1983}.

However, the presence of square hard-wall boundaries affects the system configuration even when the number of particles is very large, despite the fact that the interparticle and wall-particle interactions are only contactlike. Due to the incompatibility between the symmetry of long-range crystalline order and the symmetry of the container, random arrangements occur with greater probability than observed with PBCs. Confinement, therefore, offers an alternative route to study random packing of frictionless monodispersed disks that would otherwise crystallize into a triangular lattice.

It is interesting to observe that under the current protocol, our model retrieved the {\em optimal\/} packing\footnote{By optimal packing we mean the densest arrangement of circles in a square container; a mathematical problem of geometric optimization with a venerable history that goes back to the work of Fejes T\'oth. Optimal packings of circles in square containers, for a number of circles in the range discussed in this paper, have been found by strict mathematical arguments or numerical simulations. An excellent review can be found in Ref.~\cite{circle_packing_szabo_2007}} in all cases, with the exception of $n^2$=49.  For confined systems with few particles (i.e., $n^2$$<$49), the optimal packing coincides with the most probable packings (i.e., highest frequency peak of the PDF), whereas for $n^2$$>$49, the optimal packing is only found in the tail of the distribution. At the same time, rare-event configurations (e.g., the two low-density configurations for 9 particles in Fig.~\ref{fig:selected_configurations}(a)--\ref{fig:selected_configurations}(b)) were also found during the sampling of the configuration space.

Some of the highly symmetric configurations, differing from the most probable, have small but significant sampling frequencies, for instance, the configuration in Fig.~\ref{fig:selected_configurations}(g) appears  $\sim$6\% of the times (corresponding to the second peak from the left in Fig.~\ref{fig:pdfs_bcs}(b)). An interesting continuation to this work would be on possibilities in targeting particular configurations by the combined effect of geometrical confinement with the introduction of small yet controlled deviations of the particle shape\cite{torquato_inverse_optimization_review_2009}. This is an important question, as  experiments in the real world deal, firstly, with particles whose shapes are far from the mathematical perfection of a circle (or sphere), and secondly, with particles that are by necessity always confined within given boundaries.

We are thus able to conclude that for monodispersed particles, subject to a stochastic driving force (e.g., turbulent flow), the optimal configurations are the most probable for small confined systems ($n^2$$<$25), whereas for larger systems, optimally-packed structures are found only in the tails of the distributions. These distributions, in turn, are increasingly dominated by random-packed configurations as the size of the system increases. The existence of high-symmetry configurations (other than the $n$$\times$$n$ square arrays) with non-negligible occurrence probabilities is, therefore, a direct consequence of monodispersity and confinement. This is an interesting result for applications in devices whose functionality is dependent on the state of order, regardless of the low-yield rate for symmetric configurations observed from the present protocol\cite{shen_confinement_feature_2010,kumacheva_confinement_channels_2003,gates_monodisperse_shape_size_2001,ozin_confinement_patterning_2001,kumacheva_confinement_patterning_2002}.

\section{Summary and conclusions}
\label{sec:conclusions}
In summary, separating the boundary and finite size effects by sampling a large number of packing structures with HBC and PBC, we observe that monodisperse disks crystallize only in the $n$$\times$$n$ square configuration for very small systems, $n^2$$<$36, with the triangular crystallization more probable above a confinement limit of $n^2$=49 (i.e., random packing can be seen as a protoform for triangular lattice formation). Boundary effects, while diminishing as the number of particles increases, are still visible even at large $n$ due to the mismatch in symmetry between the particle and container. As strict confinement has been seen to produce packing configurations not found often in nature, confinement can easily be used as a parameter to tune the experimental realization of two dimensional structures within colloidal suspensions, microfluidic channels, or molecular superlattices.

We believe that our results for square containers are indicative of a general behavior of confined granular disks and that the main qualitative aspects will be observed as well in containers with different shapes. A more difficult and interesting question is whether optimal configurations of particles with anisotropic shapes, e.g., ellipses, are also the most probable packings under a 2D stochastic agitation protocol. This work also leads to several other questions, including whether a packing protocol can be tuned to increase the yield rate of nontrivial high-symmetry configurations while staying amenable to experimental realization. This includes the essential question of the threshold value above which the polydispersity mitigates the ordering provided by confinement. These questions surely warrants further investigation, and we shall pursue them in future work.

\ack
The authors are grateful to Matthias Sperl and Roland Roth for illustrative and protracted discussions at the beginning of this work, and to Alina Vlad for a critical reading of the manuscript. AT acknowledges that this research was supported by a Marie Curie European Reintegration Grant within the 7th European Community Framework Programme. Computer resources from the Texas Advanced Computing Center of The University of Texas at Austin are gratefully acknowledged.

%
% Bibliography w/BibTeX:
\section*{References}
\bibliographystyle{unsrt}
\bibliography{granular_confinement}

\end{document}